# Magnetic behavior and phase diagram of epitaxial $Er_3Fe_5O_{12}$ thin films across the compensation temperature

Satakshi Pandey[1*], Antoine Barbier[1], Anne Forget[1], Brice Sarpi[2], Francesco Maccherozzi[2], Roberto Sant[3], Nicholas B. Brookes[3], Jean-Baptiste Moussy[1], Pamella Vasconcelos Borges Pinho[3,4*]

[1] Université Paris-Saclay, CEA, CNRS, SPEC, 91191 Gif-sur-Yvette, France
[2] Diamond Light Source, Harwell Science and Innovation Campus, Didcot OX11 0DE, United Kingdom
[3] European Synchrotron Radiation Facility, F-38043 Grenoble, France
[4] Université Paris-Saclay, CEA, Service de recherche en Corrosion et Comportement des Matériaux, 91191 Gif-Sur-Yvette, France

*Corresponding author: satakshi.pandey@cea.fr; pamella.vasconcelos@cea.fr

**Abstract:** Rare-earth iron garnet ($RE_3Fe_5O_{12}$) films are promising insulating ferrimagnets. They can show low magnetic damping, perpendicular magnetic anisotropy, and ultrafast spin dynamics, which makes them ideal for spin transport applications. In this work, we investigate the interaction between the magnetic sublattices in $Er_3Fe_5O_{12}$ thin films grown by pulsed laser deposition on a $Gd_3Ga_5O_{12}$ substrate. Structural and magnetic characterization reveals high-quality single-crystal growth, with compensation temperature close to the reported bulk value (~80 K). Magnetic phase diagrams based on element-specific measurements map out the regions where ferrimagnetic, canted, and aligned phases are stable across the compensation temperature. The micromagnetic dynamics resulting from perpendicular magnetic pulse perturbation of an in-plane magnetized layer was investigated at room temperature and reveals complex configurations. These results are a key feature for modulating magnetization dynamics through the compensation phenomenon, which is essential for spin-based devices operating in a low-temperature regime.

# I. INTRODUCTION

In the dynamic landscape of modern electronics, the pursuit of faster and more energy-efficient technologies has led researchers to revisit long-known materials in order to propose innovative mechanisms beyond traditional silicon-based devices. Rare-earth iron garnets (REIGs), with general formula $RE_3Fe_5O_{12}$ (RE =Y, Tm, Er, Lu, Yb, Dy, Gd) have garnered significant attention in various research fields, including spintronics [1], magnonics [2], [3] and photonics [4], due to their insulating ferrimagnetic characteristics, small Gilbert damping parameter, high Curie temperature ($T_C > 500$ K) and strong Faraday rotation. They can generate and transmit magnons over long distances at room temperature, making them ideal for building next-generation magnetic memory and logic devices [3], [5].

Most of the desirable properties of REIGs rely on site occupation and exchange between rare-earth (RE) and iron cations within the garnet crystal lattice and their respective magnetizations. The magnetic structure of REIG films consists of three sublattices, with Fe cations distributed between octahedral (*a*) and tetrahedral (*d*) sites, while RE cations are located in dodecahedral (*c*) sites. Like other magnetic oxides (ferrites, manganites, *etc.*) the dominant coupling between different cations is the super exchange over oxygen anions at their vertices. Therefore, these couplings have different strengths for the different symmetries. The $Fe_d$-$Fe_a$ interaction is very strong and its antiferromagnetic alignment can only be broken above Curie temperature or in overly strong fields. Contrarily, the RE interaction with the net moment of Fe ($RE_c$-$Fe_{net}$) is much weaker and requires moderate applied fields to compete with molecular fields. The magnetic behavior has been adequately reproduced using a two-sublattice model [6].

Due to the system nature, REIG containing heavy rare-earth may exhibit magnetic compensation temperature ($T_{comp}$) [7], [8] at which the magnetic moments of $Fe_{net}$ and RE sublattices align in opposite directions, effectively cancelling each other out, so the total magnetization of the material goes to zero. The presence of compensation points is technologically advantageous, as it can support chiral spin textures [9], domain wall motion [10], and efficient spin-orbit torque switching [11]. However, the existence and the precise positions of the magnetic compensation points are strongly influenced not only by the nature of the rare-earth cations but also by the cation distribution among the magnetic sublattices [12], whose quantitative determination remains challenging.

Here, we focus on the study of crystalline erbium iron garnet $Er_3Fe_5O_{12}$ (ErIG) thin films. Among the various heavy RE elements that exhibit a compensation temperature when in the garnet structure, $Er^{3+}$ is of special interest. This relatively unexplored material exhibits a complex magnetic phase diagram and strong spin/lattice coupling, indicating magnetostriction effects [13]. At low temperatures, changes in the cubic lattice parameter are generally correlated with the onset of Er magnetic moment reorientation. Close to the compensation temperature ($T_{comp}$ ~ 78.8 K), one observes non-collinear magnetic structures featuring "umbrella-like" [14] arrangements of the Er magnetic moments emerging from the competition between the exchange-interaction of Er and Fe ions as well as the crystal-field anisotropy. Another exciting area of interest is generating non collinear textures, such as skyrmions, in ErIG as claimed in $Tm_3Fe_5O_{12}$ (TmIG) layers in contact with a heavy metal [15]. Although the variety of magnetic phases in ErIG combined with the well-known attractive features of iron garnets can be very interesting for applications in magneto-optics and spintronics (*i.e.,* skyrmionic, spin-orbitronic), few studies [16], [17] have been carried out on ErIG thin films. In particular, the magnetic properties of thin films, which may depend on the growth conditions, film thickness, or surface and interface effects, require detailed investigation. It has been shown that ErIG thin films exhibit strain- and growth-induced magnetic anisotropy [17] as well as spin Hall-induced anomalous Hall effect (AHE) [18], but a comprehensive study of the structural quality, magnetic phase diagram, and micromagnetic characterizations of these crystalline thin films is still lacking due to the complex nature of the compound. In this work, we provide element-specific magnetic phase diagrams of the magnetization process by acquiring dichroic spectra for external magnetic fields ranging from 9 T to 0 T at temperatures between 50 K and 100 K. These measurements reveal the field and temperature-dependent evolution of the system and enable us to determine the spin configurations and reorientations of each sublattice in the garnet structure near the compensation temperature. Hence, we identify the boundaries of the canted phase, within which the antiferromagnetic domains are reoriented, as well as the formation of a phase where the $Fe_{net}$ and Er sublattices align in parallel. In contrast to bulk materials, our results show that the Fe-sublattice magnetization in thin films reverses more gradually under low external magnetic fields, extending the stability of the canted phase over tens of kelvins. In addition, the magnetic domain structure and its ability to react to an external magnetic pulse along the magnetic hard axis has also been explored to evidence the relevance of this material for spintronics applications through its high sensitivity to an external magnetic field.

## II. EXPERIMENTAL DETAILS

Erbium iron garnet ($Er_3Fe_5O_{12}$, ErIG) thin films were epitaxially grown on (111)-oriented $Gd_3Ga_5O_{12}$ (GGG) substrates by pulsed laser deposition (PLD). The polycrystalline ErIG target was prepared by solid-state reaction from a stoichiometric mixture of $Er_2O_3$ (99.9% purity) and $Fe_2O_3$ (99.998% purity) powders sintered at 1175°C, 1350°C, and 1400°C for 7 hours, 7 hours, and 5 hours, respectively. During film growth, the ErIG target was ablated using a nanosecond laser beam (Nd:YAG laser) of 355 nm in wavelength with fluences varying between 1.5 and 0.90 J·cm$^{-2}$ and a repetition rate of 2.5 Hz. The base pressure in the PLD chamber was $4\times10^{-9}$ mbar. The deposition was carried out at an oxygen partial pressure of $2.5\times10^{-1}$ mbar and substrate temperature of 700 °C controlled by an optical pyrometer. We have grown thin films of different thicknesses (from 5 to 50 nm). Here, we describe the growth, the structure and the magnetic properties for two thicknesses: 12 nm and 34 nm.

The crystalline structure of the atomic layers was monitored in real time using a reflection high-energy electron diffraction (RHEED) gun along both the $[1\bar{1}0]$ and $[11\bar{2}]$ azimuthal directions. Following growth, the film thickness was confirmed *ex-situ* by fitting the X-ray reflectivity (XRR) curves. The crystalline phase and strain were determined by X-ray diffraction (XRD) and Reciprocal Space Mapping (RSM), using a 5-circle diffractometer (*Rigaku Smartlab XE*) equipped with a 9kW Cu rotating anode ($\lambda_{Cu\,K\alpha}$ = 1.5406 Å), an incident Ge(220)×2 monochromator, and running in the parallel beam geometry. Atomic force microscopy (AFM) images were recorded using an *Aslyum Research MFP-3D* microscope in tapping mode in order to probe the surface morphology of the films.

The magnetization at saturation, the coercive field and the compensation temperature of the ErIG thin films were obtained using a vibrating sample magnetometer (VSM) in a *Quantum Design Physical Properties Measurement System* (7 T PPMS-VSM). In-plane magnetic hysteresis loops were recorded from 50 to 300 K with a magnetic field up to 10 mT. In order to extract the magnetic signal from the deposited layer, the strong paramagnetic contribution of the GGG substrate was removed by subtracting a linear contribution from the raw data.

To explore the magnetic phase diagram of ErIG thin films across the compensation temperature, X-ray magnetic circular dichroism (XMCD) measurements were carried out on the 12 nm-thick film at the Fe $L_{2,3}$-edges and Er $M_{4,5}$-edges. The measurements were performed at the European Synchrotron Radiation Facility (ESRF), beamline ID32 [19], [20]. Spectra were acquired in the temperature range of 20 to 300 K in a geometry corresponding to an angle of 60° between the beam and the sample surface. The absorption signal was recorded in total electron yield (TEY) mode. For the field-dependent series, measurements were performed at 50, 60, 70, 80, 90, and 100 K. At each temperature, spectra were sequentially acquired starting from the highest applied magnetic field and decreasing stepwise down to zero, *i.e.*, at 9, 7, 5, 3, 1, and 0.75 T. The magnetic field was applied along the beam direction. All measurements were performed in UHV conditions with base pressure inside the chamber of about $3\times10^{-10}$ mbar.

At room temperature, magnetic domains and spin textures of the 12 nm-thick ErIG film were imaged using X-ray magnetic circular dichroism combined with X-ray photoemission electron microscopy (XMCD-XPEEM). The measurements were conducted at Diamond Light Source (beamline I06) using an Aberration-Corrected PhotoEmission Electron Microscope (AC-PEEM, *Elmitec Gmbh*) equipped with a MAPS detector (Medipix-3, *Amsterdam Scientific Instruments*). To avoid strong charge build-up, a 2 nm-thick Pt capping layer was deposited by $Ar^+$ sputtering, thin enough to preserve suitable conditions for electron spectroscopy measurements. The sample was mounted on a dedicated cartridge equipped with magnetic coils enabling *in situ* application of in-plane and out-of-plane magnetic fields up to 30 mT and 70 mT, respectively. Prior to spectromicroscopy measurements, low energy electron microscopy (LEEM) alignment was used to define the region of interest and assess the surface morphology. XMCD-XPEEM images were recorded at the maximum of Fe-$L_3$ absorption edge (709.4 eV) using circularly polarized monochromatic light. Image drift was manually corrected using the magnetic pinning centers as reference points. The measurements were performed in ultra-high vacuum (~$10^{-9}$ mbar) and in grazing incidence geometry with the incoming beam at 74° from the surface normal.

### III. RESULTS AND DISCUSSION

#### A. CRYSTALLINE GROWTH AND STRUCTURAL ANALYSIS

Stoichiometric ErIG(111) thin films were grown on single crystalline GGG(111) substrates by PLD. **Figure 1a** shows the RHEED patterns along the $[11\bar{2}]$ azimuth recorded during the growth. The diffraction patterns of both the 12 nm and 34 nm-thick ErIG films exhibit sharp streaks without any additional spots, indicating a two-dimensional growth mode and the absence of secondary phases. Following the growth, the layer thicknesses, estimated from the deposition time and laser fluence, were confirmed *ex-situ* by XRR measurements (**Figure 1b**) using Parratt's formalism. The fits are in good agreement with theoretical parameters (density, atomic structural factor, and roughness) and yield an RMS roughness below 0.5 nm for both films. AFM topographic images (see **Figure S1**) also reveal low surface roughness of the grown films, which shows a surface predominantly composed of granular structures. These results are comparable to those reported in the literature [17] for other REIG thin films grown by PLD.

To confirm the single-crystalline nature of our films, *ex-situ* specular θ-2θ XRD measurements were performed. The diffractograms can be indexed using only (*hhh*)-type reflections corresponding to the (111) orientation of the garnet structure, confirming that the ErIG layer is (111)-oriented and aligned with the substrate lattice. **Figure 1c** depicts a close-up view of the (444) peak where the well-marked Laue oscillations reflect a highly coherent growth of the ErIG layer. The film thickness deduced from the Laue oscillations agrees well with the XRR measurement, indicating crystalline coherence throughout the entire film. For the 34 nm-thick film, we observe a sharp peak near $2\theta = 51.076°$, which originates from the GGG(111) substrate and corresponds to the expected lattice parameter of 1.2379 nm [21]. Additionally, a broad peak is present around $2\theta = 50.80°$, attributed to the film. From these values, we deduce the interatomic distance plane of the 34 nm-thick ErIG thin film in the out-of-plane direction $d_{444} = 0.1795$ nm, which corresponds to the lattice parameter ($c$) of 1.2441 nm if we assume a cubic symmetry. Similarly, for the 12 nm-thick film, the (444) diffraction peak position at $2\theta = 50.52°$ gives an out-of-plane lattice spacing $d_{444} = 0.1807$ nm, which corresponds to a lattice parameter of $c = 1.2506$ nm. Therefore, the out-of-plane lattice parameter of ErIG thin films has a higher value than in the bulk ErIG sample ($c = 1.2350$ nm) [22].

Additional information on the out-of-plane but also on the in-plane lattice ($a$) parameters can be obtained from reciprocal space maps (RSM) along the (642) oblique plane. The RSM scan on the 34 nm-thick film (**Figure 1d**) yields an inter-reticular distance of $d^{\text{ErIG, RSM}} = d^{\text{GGG, RSM}} = (q_x/2\pi)^{-1} = 0.438$ nm, where $q$ is the wave vector corresponding to an in-plane lattice parameter of $a = 1.2411$ nm, close to the value of GGG [23]. The inter-reticular distances in the out-of-plane direction $d^{\text{ErIG, RSM}}_{444} = (q_z/2\pi)^{-1}$ is 0.1795 nm, consistent with the lattice constants measured by $\theta$-$2\theta$ XRD. In contrast, the film peak could not be resolved by RSM for the 12 nm-thick film. It is important to highlight that the out-of-plane lattice parameter for the film is slightly larger than that of the bulk ErIG and the GGG substrate. This discrepancy may be attributed to a too low oxygen partial pressure during growth, which could lead to the formation of oxygen vacancies [24].

In summary, diffraction measurements reveal that our ErIG films are grown under significant epitaxial strain, with in-plane lattice parameters closely matching those of the GGG substrate. Conversely, the out-of-plane lattice parameter is slightly elongated, indicating tensile strain along the [111] direction. The degree of this distortion was quantified through the calculation of the distortion angle (see *Supporting Information*), yielding a value of approximately 89.90° for 34 nm films. This confirms that the films experience out-of-plane tensile strain, which is particularly relevant since the magnetic compensation temperature of rare-earth iron garnet thin films can be tuned by the substrate-induced misfit strain [25].

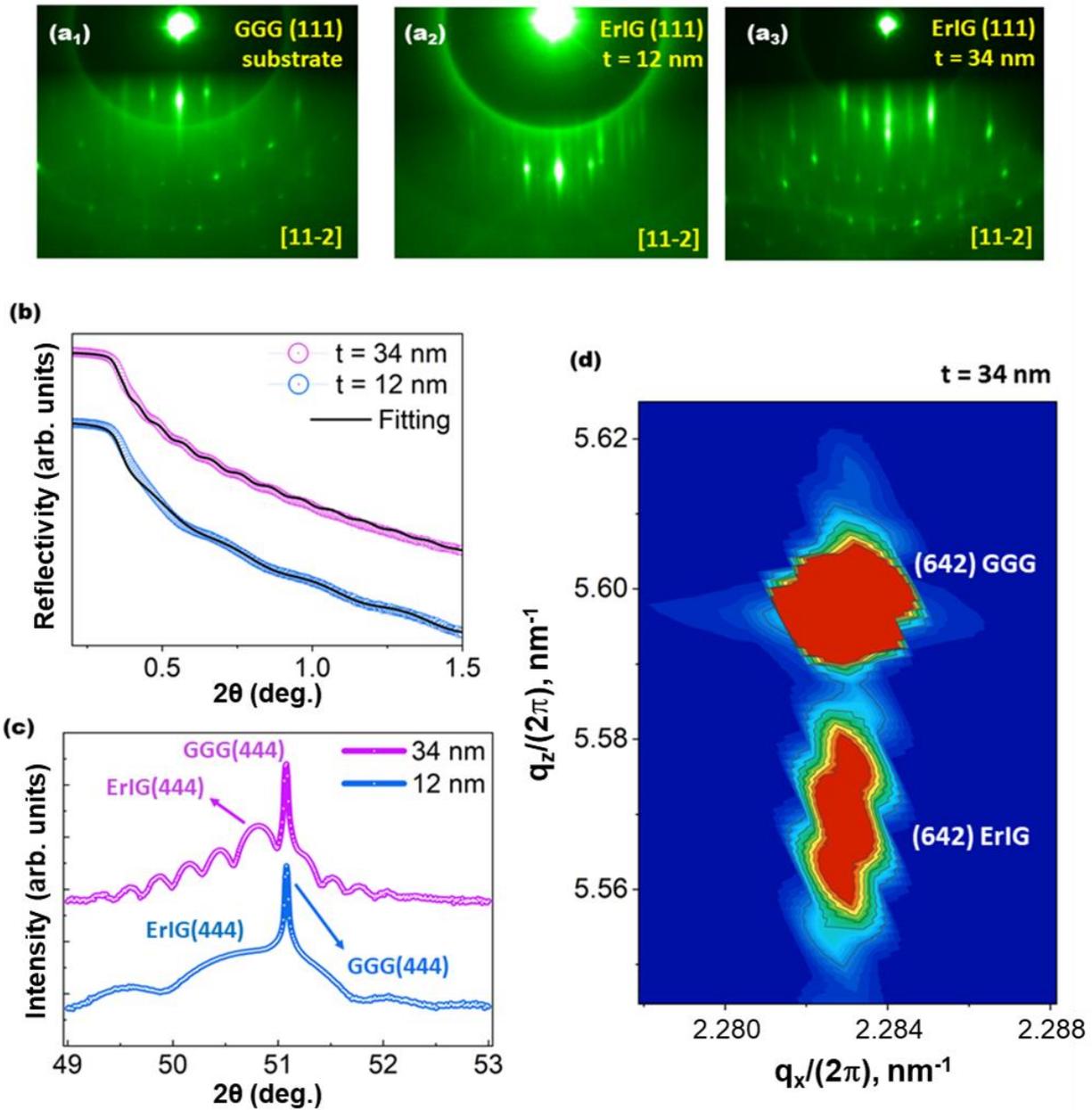

**Figure 1.** Structural characterization of ErIG thin films grown on GGG(111) substrates. RHEED patterns along the [11$\bar{2}$] azimuth of the GGG(111) substrate (a$_1$) and the deposited 12 nm (a$_2$) and 34 nm (a$_3$) ErIG(111) thin films. (b) XRR curves of 34 nm (purple) and 12 nm (blue) films with fittings using Parratt's formalism. (c) θ-2θ XRD patterns of the 34 nm (purple) and 12 nm (blue) films, showing the (444) Bragg peaks with clear Laue oscillations. (d) RSM patterns around the (642) asymmetric diffraction peak for 34 nm-thick layer.

## B. MAGNETOMETRY MEASUREMENTS

Once probed the crystalline structure of the 12 and 34 nm-thick ErIG(111) thin films, we investigated their macroscopic magnetic properties.

**Figure 2a** shows a schematic view of ErIG cationic arrangements where $Fe^{3+}$ cations sit at tetrahedral (d) and octahedral (a) sites, while $Er^{3+}$ cations sit at dodecahedral (c) sites. At temperatures above the compensation temperature ($T > T_{comp}$), the super-exchange interaction between Fe sublattices dominates, while at temperatures $T < T_{comp}$, Er sublattice contributes significantly owing to its strong spin-orbit coupling. Around $T_{comp}$, the three sublattices are non-collinear and give rise to a canted phase [26]. The total magnetization is then the sum of the spin contributions from each sublattice ($Fe_d$, $Fe_a$ and $Er_c$), accounting for their antiferromagnetic coupling. Here, the total magnetization of the ErIG thin films across $T_{comp}$ is probed using standard VSM magnetometry. **Figure 2b** shows the in-plane (IP) and out-of-plane (OOP) magnetic hysteresis loops of the 34 nm-thick ErIG film measured at room temperature. The loops reveal an in-plane easy axis, with a saturation field of approximately 0.2 mT and a magnetic coercivity ($H_C$) of about 0.1 mT (see **Figure S2a** for a zoomed-in view of the IP loop), confirming the soft magnetic nature of erbium iron garnets. The remanent magnetization ($M_R$) corresponds to 80% of the magnetization at 0.5 mT and the saturation magnetization ($M_S$) is around 50 kA·m$^{-1}$. Similar to the thicker film, the 12 nm film also exhibits an in-plane easy axis (see **Figure S3**). However, its $M_S$ is lower (about 15 kA·m$^{-1}$ at room temperature), as shown in **Figure S2b** (dark red line). Notably, the $M_S$ values obtained for our films are lower than those typically reported in the literature for ErIG(111) thin films grown on other garnet substrates with perpendicular magnetic anisotropy (95 kA·m$^{-1}$ [18]). This reduction in magnetization likely comes from slight deviations in cationic stoichiometry or oxygen content [27].

To probe the existence of $T_{comp}$ in thin layers and their magnetic behavior across it, IP magnetic hysteresis loops were recorded at different temperatures. **Figure S2** shows the IP loops of the 34 nm and 12 nm films after subtraction of the strong paramagnetic contribution from the GGG substrate. **Figure 2c (2e)** depicts the temperature dependence of $M_S$, while **Figure 2d (2f)** displays the temperature dependence of $\mu_0 H_C$ for the 34 nm (12 nm) films, as extracted from the IP loops shown in **Figure S2**. We clearly observe a divergence in the coercive field with a peak around 80 K, which is the signature of the temperature of compensation. Near the compensation point, there is a strong interaction between Er and Fe sublattices due to the increased antiparallel alignment of magnetic moments. Hence, a peak is observed in $\mu_0 H_C$ at this temperature. The enhanced antiferromagnetic coupling between the cation sublattices near $T_{comp}$ leads to a reduction in the net magnetization, as the magnetic moments of the two sublattices become nearly balanced. Below $T_{comp}$, the magnetization increases rapidly. A pronounced dip in the magnetization near the compensation point is clearly observed for the 34 nm-thick film. In contrast, the transition is less distinct in the thinner film, most likely because the strong paramagnetic contribution from the GGG substrate masks the signal from the thin layer. The temperature dependence of $\mu_0 H_C$, $M_S$, anf $M_R$ highlights the crucial role of the exchange interactions in governing the magnetic behavior of the ErIG thin films.

In the following sections, we focus on the magnetic phase diagram and micromagnetic dynamics of the 12 nm-thick film. These measurements were performed using surface-sensitive, element-specific techniques that are not biased by substrate contributions, enabling a reliable investigation of the magnetic properties of thinner films.

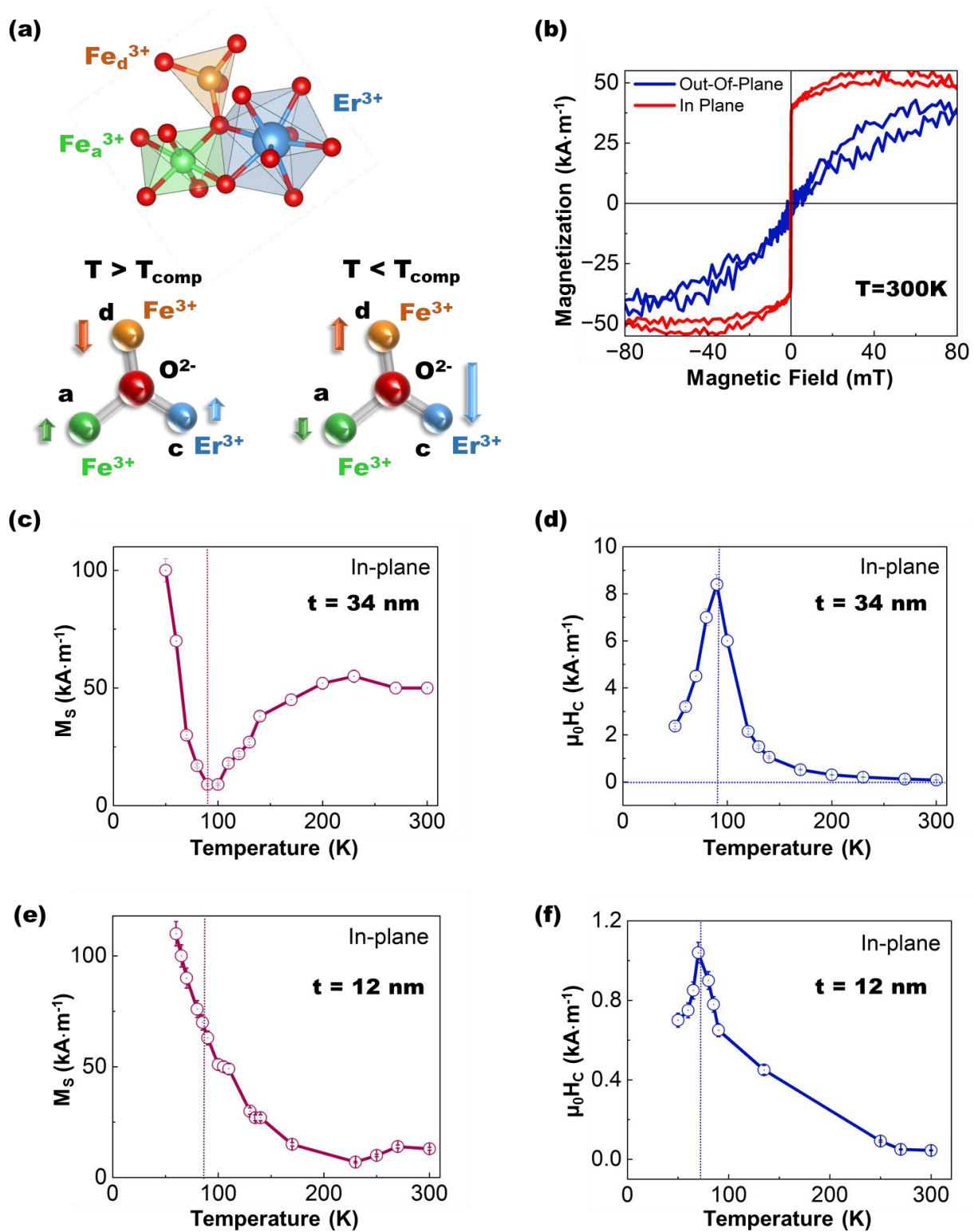

**Figure 2.** Magnetic characterization of ErIG thin films grown on GGG(111) substrates. (a) Schematic representation of the three magnetic sublattices for temperatures above and below the compensation temperature. The labels a, d, and c indicate crystallographic symmetries: tetrahedral (Fe ions), octahedral (Fe ions), and dodecahedral (Er ions), respectively. (b) Magnetic hysteresis loops measured by VSM at room temperature, showing that the 34 nm-thick ErIG film exhibits in-plane magnetic anisotropy. (c, e) Saturation magnetization ($M_S$) and (d, f) coercive field ($\mu_0 H_C$) as a function of temperature for the 34 nm and 12 nm films, respectively.

## C. MAGNETIC PHASE DIAGRAM

Element specific investigations allow providing additional insights in the mechanisms behind the occurrence of $T_{comp}$. An element-specific description of the magnetic behavior of the 12 nm-thick film across $T_{comp}$ was obtained using XMCD spectroscopy. **Figure 3** shows Fe $L_{2,3}$-edges and Er $M_{4,5}$-edges XAS and XMCD spectra recorded at room temperature and external magnetic field of 0.25 T. XMCD spectra were obtained as the difference between XAS spectra recorded with right- and left- circularly polarized beam at a given absorption edge. This technique is site- and element-specific and is sensitive to the magnetic polarization of each species.

The Fe $L_{2,3}$-edges XAS and XMCD spectra (**Figure 3a**) are divided in the spin-orbit split $L_3$ (~709 eV) and $L_2$ (~722 eV) parts. A weak feature is also visible at 701.7 eV, which corresponds to the second-order harmonic of the Er absorption edge. The Fe-$L_3$ dichroic signal shows three main peaks centered at 708.1 eV, 709.3 eV and 709.9 eV. According to crystal field multiplet calculations [28] (see **Figure S4**), the first contribution corresponds to $Fe_a^{3+}$ cations, the second to $Fe_d^{3+}$ cations and the third to a mixture of these two sites. The Er $M_{4,5}$-edges XAS and XMCD spectra (**Figure 3b**) are also divided in the spin-orbit split $M_5$ (~1403 eV) and $M_4$ (~1445 eV) parts and display the multiplet structure typical of the ionic nature of erbium. In this case, Er-$M_5$ dichroic signal has only a main contribution centered at 1403.5 eV, which corresponds to $Er_c^{3+}$ cations. It is noted that the dichroic signals from $Fe_d^{3+}$ and $Fe_a^{3+}$ cations, as well as $Fe_d^{3+}$ and $Er_c^{3+}$ cations, have opposite signs, indicating that these species are coupled antiferromagnetically, consistent with the literature [26].

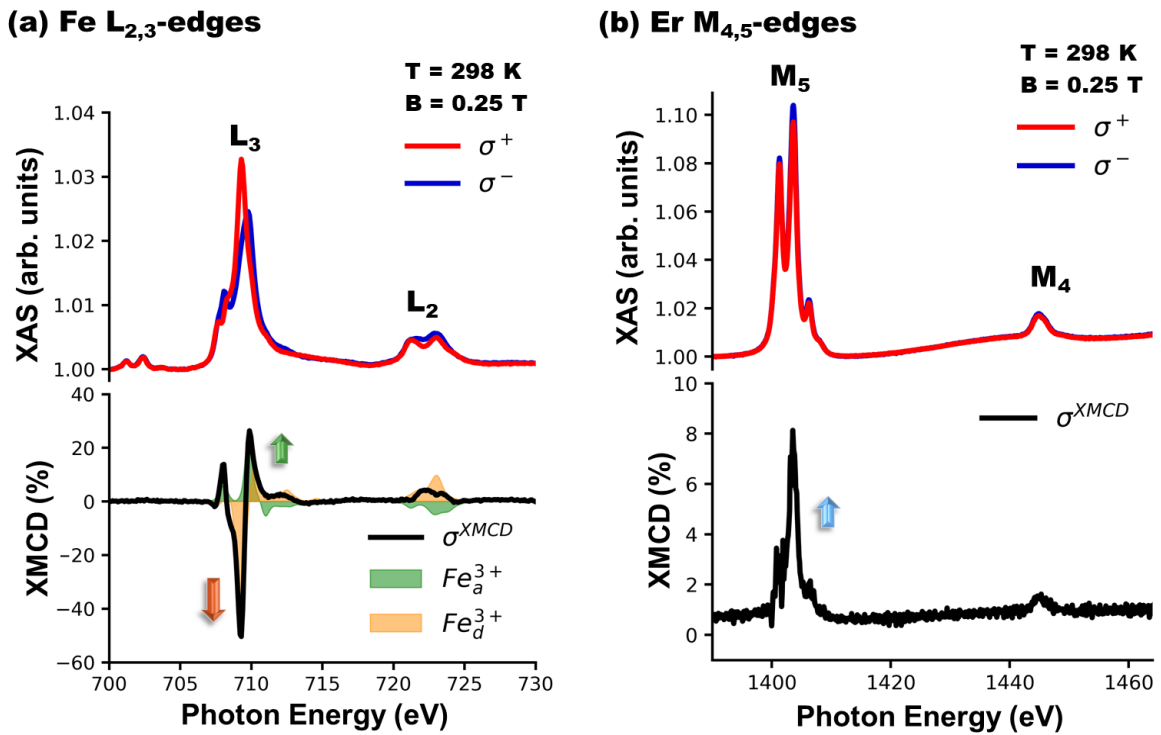

**Figure 3.** In-plane XAS spectra (top) at the (a) Fe $L_{2,3}$-edges and (b) Er $M_{4,5}$-edges for right ($\sigma^-$) and left ($\sigma^+$) circular polarization, together with the corresponding difference spectra ($\sigma^{XMCD}$, bottom). Measurements were performed at room temperature under an external magnetic field of 0.25 T for the 12 nm-thick ErIG film. The calculated contributions for each iron species are shown as $Fe_a^{3+}$ (green) and $Fe_d^{3+}$ (orange).

The ferromagnetic ordering in ErIG therefore arises from the strong antiferromagnetic coupling between an inequivalent number of $Fe^{3+}_d$ and $Fe^{3+}_a$ cations (3:2 per formula unit) and the $Er^{3+}_c$ cations, the latter being coupled antiferromagnetically to $Fe_{net}$. Unlike the Fe sublattices, for which the antiferromagnetic alignment ($\uparrow_d\uparrow_d\uparrow_d\downarrow_a\downarrow_a$) can be broken only above Curie temperature (544 K), the Er sublattice moment is more sensitive to temperature and moderate external fields. According to the previous VSM measurements (**Figure 2**), $T_{comp}$ for the ErIG thin films is approximately 80 K. To further investigate the magnetic interaction between the Er and Fe sublattices, **Figure 4** compares the XMCD spectra recorded for temperatures around $T_{comp}$ (*i.e.*, 70 K, 80 K and 90 K) under applied fields ranging from 9 T to 0.25 T. Here, we clearly observe the characteristic peaks of $Fe^{3+}_d$, $Fe^{3+}_a$, and $Er_c^{3+}$ cations and their antiferromagnetic coupling, as described above. Three effects are evident in the dichroic signal as a function of temperature and magnetic field: (*i*) the amplitude of the Er dichroic signal increases for T < $T_{comp}$, (*ii*) the relative sign of the Fe and Er dichroic signals reverses when crossing $T_{comp}$, and (*iii*) the dichroic signal vanishes at specific temperature and field values (*e.g.*, at T = 80 K and B = 1 T). The latter behavior arises from the ability of two-sublattice ferrimagnets to form a canted phase below a critical temperature ($T_{crit}$), through which the sublattice magnetizations can be reversed [26], [29]. To have a comprehensive understanding over the magnetization process, including the inversion of the individual components and the strength of their interactions, we construct a magnetic phase diagram (**Figure 5**) using the amplitude of the XMCD signals for the Fe and Er sublattices in the range 50-100 K and 9-0.25 T. In this diagram, the arrows represent the orientation of the sublattice magnetization vectors: $Fe_{net}$ (**Figure 5b**) and Er (**Figure 5b**). The XMCD amplitude and sign are proportional to the angle between each sublattice moment and the direction of the applied magnetic field ($B_{ext}$). Arrows pointing upward or downward correspond to XMCD maxima, indicating that the sublattice moments are saturated and aligned parallel or antiparallel to the external field. In contrast, horizontal arrows represent XMCD values near zero, consistent with sublattice moments oriented perpendicular to the external field.

According to **Figure 5**, the phase diagram of ErIG comprises a canted phase (C) between 70 and 90 K that separates two ferrimagnetic regions (F1 and F2), in which the $Er^{3+}_c$ cations are coupled antiferromagnetically to $Fe_{net}$. **Figure 5a** shows that the coupling between the $Fe^{3+}_d$ and $Fe^{3+}_a$ sites is so strong that the Fe net magnetization is independent of field and temperature outside the canted phase, and it reverses its orientation by 180° within the canted phase through continuous rotation. The Er-sublattice (**Figure 5b**), on the other hand, is weakly coupled to the Fe-sublattice, so their interaction can be broken at moderate fields. The Er-sublattice behaves like a paramagnet outside the upper boundary of the canted phase, and its magnetization is cancelled by an applied field of around 6 Tesla (solid black line). Above this field, one observes the reversal of Er-sublattice magnetization and a phase (A) appears in which the Er-sublattice is parallel aligned to the net Fe moment and the external magnetic field. Below the compensation temperature, one evidences the deflection of the Er moments away from the direction of the applied field simultaneous to the rotation of the Fe-sublattice in the canted phase. This phase diagram of ErIG(111) thin films is slightly different from that predicted for the bulk ErIG(001) structure using an isotropic two sublattice model [26]. Although we find the same phases (F1, F2, C and A) at similar temperatures and fields, the extent of the canted phase is larger for the thin films than expected for the bulk material. Indeed, in the thin films, the Fe and Er sublattice magnetizations are reversed more progressively at low external magnetic fields, extending the canted phase to tens of Kelvins.

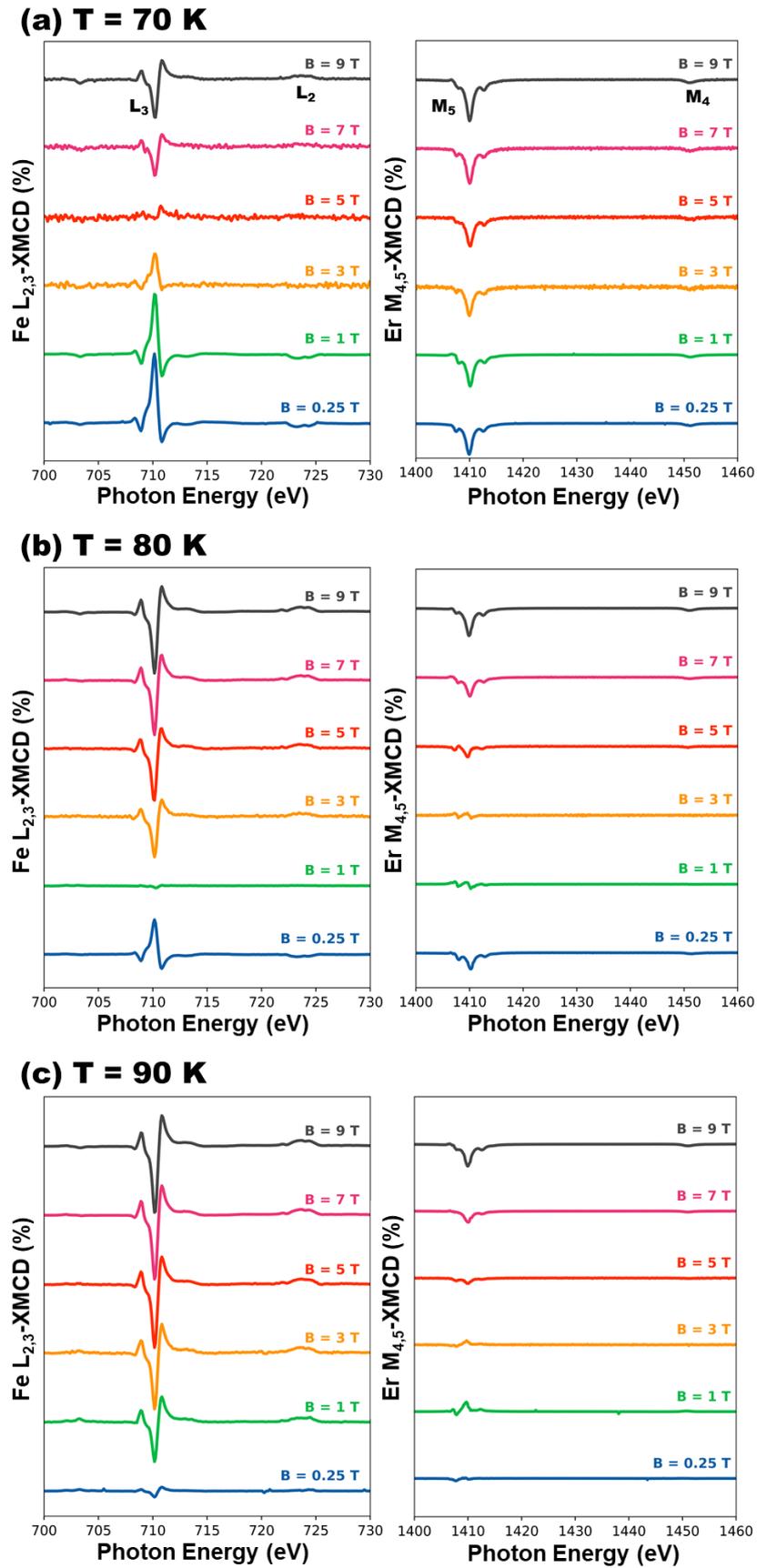

**Figure 4.** In-plane XMCD spectra for 12 nm-thick ErIG at the Fe $L_{2,3}$-edges (right) and Er $M_{4,5}$-edges (left) measured at (a) T = 70 K, (b) T = 80 K and (c) T = 90 K under applied fields ranging from 9 T to 0.25 T.

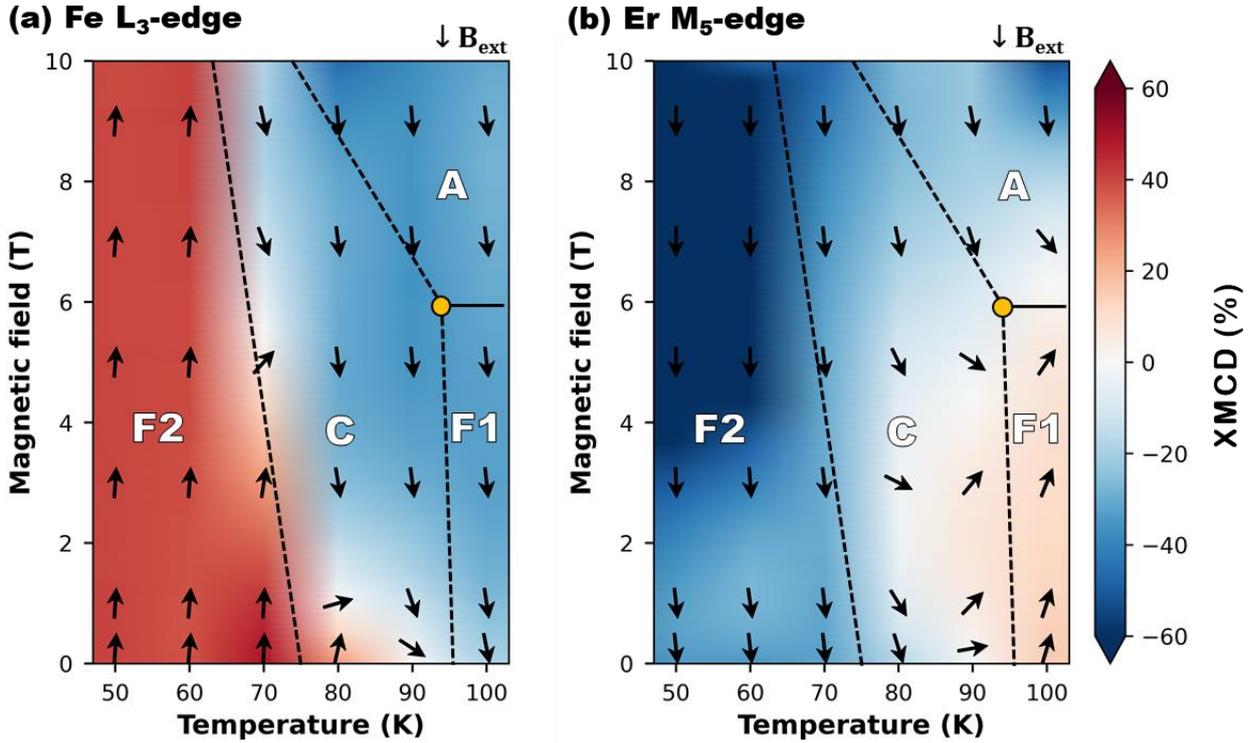

**Figure 5.** Magnetic phase diagram of ErIG(111) thin films obtained from the interpolation of (a) Fe L$_3$-edge and (b) Er M$_5$-edge XMCD amplitude with the magnetic field applied in-plane. The diagram includes two ferrimagnetic regions (F1 and F2), a canted phase (C), and a region where the two sublattices are aligned (A). Arrows indicate the spin orientations of each sublattice relative to the external magnetic field (↓B$_{ext}$) and are placed at the corresponding XMCD data points. Dashed and solid black lines serve as guides to the eye, estimating the lower and upper limits of the canted phase and the point at which the Er sublattice demagnetizes, respectively. The yellow circle marks the critical temperature ($T_{crit}$). Error bars are ± 1K for temperature and ± 0.005 T for external magnetic field.

**D. MICRO-MAGNETIC PROBE AT ROOM TEMPERATURE**

The magnetic domain structure and response to external magnetic fields of the 12 nm-thick ErIG layer capped with a 2 nm-thick conductive layer of platinum (Pt) was investigated at room temperature using XPEEM spectromicroscopy. Note that the presence of the Pt overlayer could induce Dzyaloshinski-Moriya (DMI) interactions at the interface promoting the presence of non collinear textures [30], [31], [32]. Prior to the XPEEM investigations, the ErIG(12 nm)/Pt(2 nm) bilayer was magnetized under a 30 mT in-plane field to align most magnetic domains along the easy axis. Subsequent *in situ* application of magnetic pulses enabled direct observation of the evolution of the magnetic domain structure. **Figure 6a** shows an XMCD-XPEEM image at Fe L$_3$-edge (**Figure S5**) of the magnetic domain structure prior to magnetic pulses. We observe a region of the sample exhibiting varied magnetic behaviors within an otherwise predominantly monodomain film. The sample was then exposed to a succession of out-of-plane magnetic pulses having a duration of 10 s and values of 0.36, 1.1, 1.8, 3.6, 5.5, 7.3, 11, 14, 18, 25, 33, and 46 mT. Images were acquired at remanence; **Figures 6b-d** show different resulting magnetic configurations after a pulse.

**Figure 6a** highlights seven regions with distinct magnetic behaviors, labeled A to G. No changes were observed up to a pulse value of 5.5 mT, and below this threshold, the system always returned to its initial remanent state after the pulse. Above this value (**Figures 6b-d**, full sequence in **Figure S6**), the different regions exhibited varied responses, as detailed in the *Supporting Information*. For instance, Region A corresponds to a large single domain that switches its orientation completely at 7 mT and remains stable under higher fields. In contrast, Region C shows no in-plane magnetic response, most likely due to a perpendicular magnetic orientation stable for all pulse values. Region D adopts also a surprisingly different behavior since it switches from the pinned reverted orientation (opposed to Region A) to an out-of-plane orientation with moderate field values (~15 mT). The remaining regions are located near pinning centers; Region E switches only at higher fields, while Regions F and G remain fully pinned and unaffected by any applied pulses. The intensity of XMCD signal was extracted for each region as reported in **Figure 6e**. While most regions exhibit the expected in-plane magnetic reversal under the applied out-of-plane magnetic pulses, regions C and D display persistent perpendicular magnetic orientations, which could be consistent with the presence of topological objects such as skyrmions.

In addition, the domain walls and magnetization orientations were analyzed after applying a sequence of positive and negative magnetic pulses to allow the magnetic configuration to reach a steady state (see **Figure S7** and the *Supporting Information* for details). The domain walls are identified as Néel-type, exhibiting in-plane rotation of the magnetic moments. These results suggest that, although the film is largely magnetically homogeneous, its domain structure and dynamics are influenced by pinning centers, exchange interactions, and localized topological effects, which may be relevant for future spintronic applications.

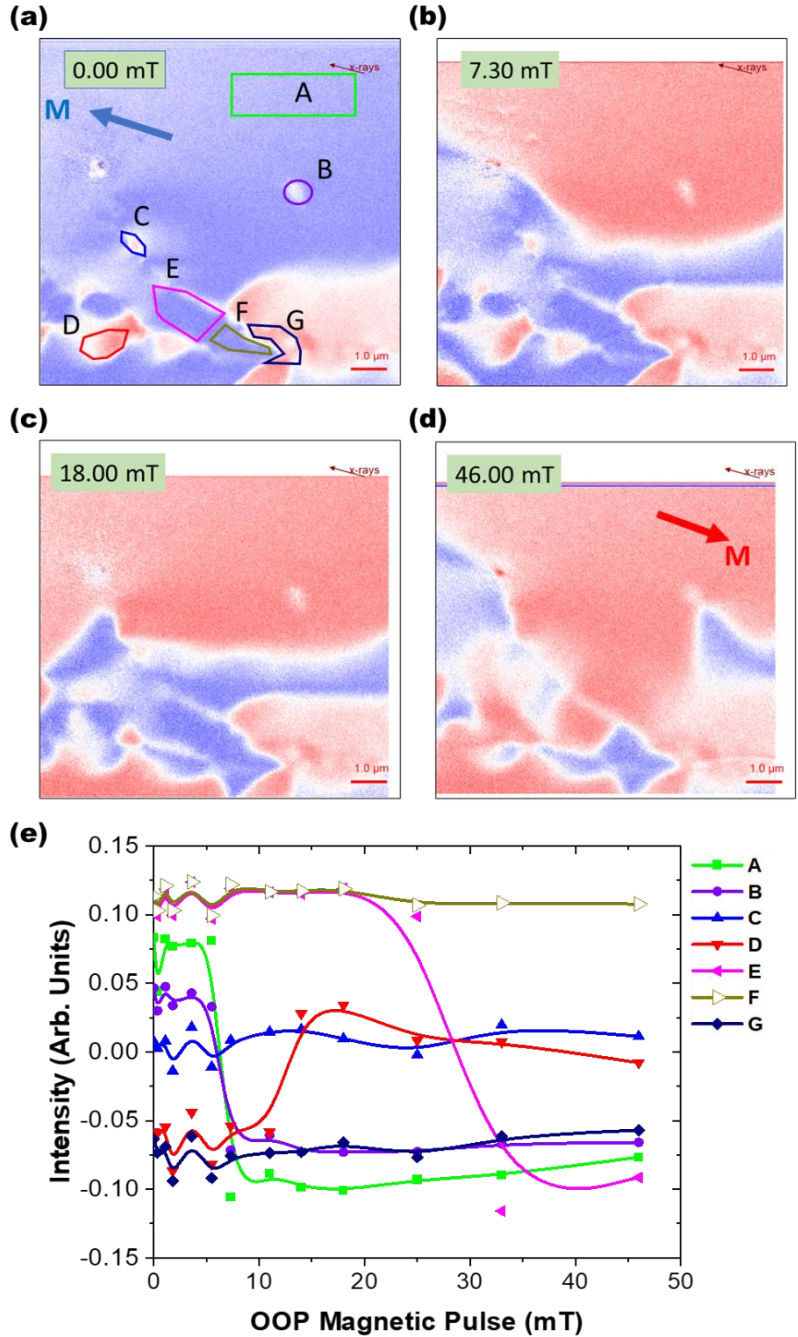

**Figure 6**. Out-of-plane magnetic response of an in-plane magnetized ErIG(12 nm)/Pt(2 nm) bilayer. (a) XMCD-XPEEM image at Fe $L_3$-edge showing the initial domains structure prior to application of out-of-plane magnetic pulses. (b-d) Domain configuration after applying different out-of-plane magnetic pulses. (e) Integrated XMCD intensity of the domains indicated in (a) with the same color code and lettering, the points are connected by spline lines.

## V. CONCLUSIONS

This work provides a detailed study of the structural quality, magnetic behavior, and phase diagram of ErIG(111) epitaxial thin films. It further highlights the temperature dependence of the exchange interaction, which gives rise to a compensation temperature. Our results demonstrate that crystalline ErIG thin films provide a versatile platform to modulate the stability ranges of distinct magnetic phases. By altering the relative contributions of the Er and Fe sublattices, we can shift the existence domains of ferrimagnetic, canted, and aligned phases. In addition, the in plane magnetic domains configuration was investigated for the 12 nm-thick film. It shows Néel walls as well as topological objects of interest with perpendicular magnetization that are currently sought in spintronics. Both the presence of compensation temperature and magnetic domains are particularly relevant for the fabrication of spin-based devices.

## ACKNOWLEDGMENTS

The authors acknowledge the ANR-FRANCE for its financial support of the DeMIuRGe project n°: ANR-22-CE30-0014. The authors would like to thank the ESRF and Diamond Light Source for providing beamtime under the project numbers IH-HC-3814 (doi:10.15151/ESRF-ES-946918332) and MM-31889, respectively. The authors also would like to thank the staff of beamline ID32 (ESRF) and I06 (DIAMOND) for assistance with data collection and providing excellent working conditions. Dr. Yuran Niu (MAX-IV synchrotron, Lund, Sweden) is also acknowledged for lending the magnetic cartridge used for the XPEEM measurements.